\documentstyle[prl,aps,amsfonts,twocolumn]{revtex}

\newcommand{\bea}{\begin{eqnarray}} 
\newcommand{\eea}{\end{eqnarray}}
\newcommand{\beann}{\begin{eqnarray*}} 
\newcommand{\eeann}{\end{eqnarray*}}
\newcommand{\beq}{\begin{equation}} 
\newcommand{\eeq}{\end{equation}}

\begin{document}

\title{
{\normalsize\begin{flushright}
MPI-MIS-56/2000\\
hep-th/0009133\\[1.5ex]
\end{flushright}}
Hidden symmetries of supersymmetric $p$-form gauge theories
}

\author{
Friedemann Brandt
}

\address{ 
Max-Planck-Institute for Mathematics in the Sciences,
Inselstra\ss e 22-26, D-04103 Leipzig, Germany
\\[1.5ex]
\begin{minipage}{14cm}\rm\quad
Field theories with $p$-form gauge potentials 
can possess ``hidden'' symmetries leaving the field strengths 
invariant on-shell without being gauge symmetries on-shell.
The relevance of such symmetries to supersymmetric models
is discussed. They provide central charges of
supersymmetry algebras, play a particular r\^ole in
duality relations, and lead to peculiar interactions. 
A multiplet of N=2 supersymmetry in four
dimensions with two hidden central charges is presented.
\\[1ex]     
PACS numbers: 11.30.Pb, 11.30.-j, 11.15.-q
\\
Keywords: p-form gauge fields, supersymmetry, 
conservation laws, central charges, duality        
\end{minipage}
}

\maketitle

This letter is devoted to a particular type of symmetries
that field theories with $p$-form gauge potentials can have.
These symmetries leave the field strengths invariant on-shell but
are not gauge symmetries on-shell. 
For this reason they will be called ``hidden'' symmetries here.
We shall first determine all such symmetries for Maxwell-type
actions using results on the so-called 
characteristic cohomology of the field equations
derived in \cite{HKS}.

The remainder of the letter will focus
on the relevance of hidden symmetries
in the context of global supersymmetry. A particular aspect
is that hidden symmetries can occur in the commutators
of global supersymmetry transformations and then
give ``hidden central charges'' of the supersymmetry algebra.
This makes them play a particular r\^ole
in dualities relating supersymmetry multiplets, among others.
Another important aspect is the
relevance of hidden symmetries to the construction of 
consistent supersymmetric interactions in theories with
$p$-form gauge potentials.

A prominent supersymmetry multiplet with a
hidden central charge symmetry is the
vector-tensor (VT) multiplet of $N=2$ supersymmetry
in four-dimensional spacetime \cite{VT} (see also \cite{dWKLL}).
Another example, a vector-tensor-tensor (VTT) multiplet
with two hidden central charge symmetries,
will be presented below.

The VTT multiplet illustrates also a related feature that
will be discussed here and has
been already observed in \cite{TT} when analyzing
the $N=2$ double tensor multiplet:
the on-shell commutator of two supersymmetry transformations
may contain gauge transformations which
involve explicitly the spacetime coordinates $x^\mu$, even when
the supersymmetry transformations themselves do not depend explicitly
on the $x^\mu$.

\section*{Characteristic cohomology}

The hidden symmetries to be discussed are closely related
to the so-called characteristic cohomology of the field equations
which may also be called 
the ``cohomology of the exterior derivative
$d=dx^\mu \partial_\mu$ on-shell''. 
We shall only sketch the basic concept here.
The precise definition is made
in so-called jet spaces whose coordinates are the
spacetime coordinates and the fields and their derivatives, 
see e.g.\ \cite{report} and references therein. 
  
The cocycles of the 
characteristic cohomology are local 
$p$-forms $\omega_p$ (i.e., differential forms on some finite dimensional
jet space) which are $d$-closed on-shell. This is denoted
\beq
d\omega_p\approx 0,
\label{coc}
\eeq
where $\approx$ denotes equality on-shell. In a Lagrangean field
theory the field equations read 
$\hat\partial L/\hat\partial\phi^i=0$ where 
$\hat\partial L/\hat\partial\phi^i$ are
the Euler-Lagrange derivatives of
the Lagrangian with respect to the
fields. By definition, equality on-shell is then equality modulo
a combination of these Euler-Lagrange derivatives and
derivatives thereof,
\beq
X\approx Y\ :\Leftrightarrow\ X-Y=\sum_{k\geq 0}
P^{i\mu_1\dots\mu_k}
\partial_{\mu_1}\dots \partial_{\mu_k}
\frac {\hat \partial L}{\hat \partial\phi^i}\ ,
\label{approx}
\eeq
where $X$, $Y$ and the $P^{i\mu_1\dots\mu_k}$ are local 
forms of the fields and the range of the summation index $k$ is finite.

A cocycle of the characteristic cohomology is called trivial
(a coboundary) if it is $d$-exact on-shell,
\beq
\omega_p\approx d\omega_{p-1}\ .
\label{cob}
\eeq

The cocycles with form-degree 
$p=n-1$ (in $n$-dimensional spacetime) are conserved currents written
as differential forms; the representatives with lower non-zero
form-degree are sometimes called ``higher order conservation laws''.
One can prove on fairly general assumptions, that the
characteristic cohomology is locally trivial at all non-zero form-degrees
$p<n-1$ for theories without nontrivial gauge symmetry.
However, in gauge theories it may be nontrivial also at lower
form-degrees. The lowest possible non-zero form-degree
at which it can be nontrivial is then related to the reducibility
order of the gauge symmetry \cite{BBH1}. 

\section*{Hidden symmetries}

Let us first discuss purely bosonic actions
of the Maxwell type,
\bea
S&=&\frac 12\int d^nx \sum_a\frac{(-)^{p_a}}{(p_a+1)!}\,
\sqrt{g}\, F^a_{\mu_0\dots\mu_{p_a}}F^{a \mu_0\dots\mu_{p_a}}
\nonumber\\
&=&\frac 12\int \sum_a (-)^{np_a}
dA^a\wedge\star dA^a.
\label{L}
\eea
$A^a$ are $p_a$-form gauge potentials with possibly
different degrees $p_a\in\{1,\dots,n-2\}$, 
$F^a_{\mu_0\dots\mu_{p_a}}$ are the
corresponding field strengths and $\star$ denotes
Hodge dualization, using
\beann
A^a&=&\frac 1{p_a!}\,dx^{\mu_1}\wedge\dots \wedge dx^{\mu_{p_a}}
A^a_{\mu_1\dots\mu_{p_a}}\ ,
\\
dA^a&=&\frac 1{(p_a+1)!}\,dx^{\mu_0}\wedge\dots \wedge dx^{\mu_{p_a}}
F^a_{\mu_0\dots\mu_{p_a}}\ .
\eeann
The spacetime metric $g_{\mu\nu}$ which occurs in the action
is supposed to be a fixed background metric with
Lorentzian signature $(+1,-1,\dots,-1)$.
The action is invariant under the gauge transformations
\bea
&&\delta_{\mathrm{gauge}}(\epsilon)A^a=d\epsilon^a\ ,
\nonumber\\
&&\epsilon^a=\frac 1{(p_a-1)!}\,
dx^{\mu_1}\wedge\dots \wedge dx^{\mu_{p_a-1}}
\epsilon^a_{\mu_1\dots\mu_{p_a-1}}
\label{gsymm}
\eea
where $\epsilon^a_{\mu_1\dots\mu_{p_a-1}}$
are arbitrary gauge parameter fields.

Let us now look for the hidden symmetries of the action.
According to our definition, they are generated by transformations
which vanish on-shell on the field strengths,
\beq
\delta_{\mathrm{hidden}} F^a_{\mu_0\dots\mu_{p_a}}\approx 0\
\Leftrightarrow\ d(\delta_{\mathrm{hidden}} A^a)\approx 0.
\label{Delta}
\eeq
Hence,
$\delta_{\mathrm{hidden}} A^a$ is $d$-closed on-shell.
Furthermore it is not $d$-exact on-shell because otherwise
$\delta_{\mathrm{hidden}} A^a$ were a 
particular gauge transformation on-shell. Hence, 
$\delta_{\mathrm{hidden}} A^a$
is a nontrivial cocycle of the characteristic
cohomology of the field equations.

The relevant cohomology groups have been computed
in a flat background in \cite{HKS}.
It can be checked that the computation in \cite{HKS} goes through 
also in a general background.
One obtains that the characteristic
cohomology is represented (locally) at all form-degrees $0<p<n-1$
by the linearly independent exterior
products of the Hodge dualized field strength forms
$\star dA^a$ which can be built at the respective
form-degree. Hence, $\delta_{\mathrm{hidden}} A^a$ 
is a linear combination of such exterior products with
form-degree $p_a$. In addition,
$\delta_{\mathrm{hidden}}$ must be a symmetry, 
i.e., it must leave the Lagrangian invariant
modulo a total derivative. 
This is equivalent to the requirement that 
the Euler-Lagrange derivative of $\delta_{\mathrm{hidden}} L$
with respect to each field $A^a_{\mu_1\dots\mu_{p_a}}$
must vanish and gives
\beq
\delta_{\mathrm{hidden}} A^a=\sum
c^{aa_1\dots a_r}
(\star dA^{a_1})\wedge\dots\wedge(\star dA^{a_r})
\label{h1}
\eeq
where the sum runs over all sets $\{a_1,\dots,a_r:r=1,2,\dots\}$
such that the form-degrees of the left and right hand
sides in eq.\ (\ref{h1}) match,
\beq 
\sum_{i=1}^r(n-p_{a_i}-1)=p_a\ ,
\label{h1a}\eeq
and the $c^{aa_1\dots a_r}$ are constant
coefficients with the following symmetry properties,
\bea
&&c^{a_0\dots a_i a_{i+1}\dots a_r}=(-)^{(n-p_{a_i}-1)(n-p_{a_{i+1}}-1)}
c^{a_0\dots a_{i+1} a_i\dots a_r}
\nonumber\\
&&(\forall \, a_i,i=0,\dots,r-1).
\label{h2}
\eea

Eqs.\ (\ref{h1}) through (\ref{h2}) provide all
hidden symmetries of an action (\ref{L}). 
The corresponding Noether currents, written as
local $(n-1)$-forms, are
$j_{n-1}=\sum c^{a_0\dots a_r}
(\star dA^{a_0})\wedge\dots\wedge(\star dA^{a_r})$.

Suppose now that (\ref{L}) is only one part of an action whose
other part does not contain the fields $A^a$
but only additional (``matter'') fields
which do not bring in nontrivial gauge symmetries
(i.e., the nontrivial gauge symmetries of the full action
are still exhausted by (\ref{gsymm})).
Using the methods established in \cite{BBH1,HKS}
one can then show on fairly general assumptions that
the result on the characteristic cohomology
described above remains valid, i.e., the
matter fields give no contributions to the characteristic
cohomology at form-degrees $<n-1$, at least locally (however,
they may contribute at form-degree $n-1$). In particular, this
holds when the action is quadratic in the matter fields.
More generally, it holds for
actions satisfying an appropriate ``normality condition'', see
\cite{BBH1,BBH2,report}. As an immediate consequence, the hidden 
symmetries of such actions are still exhausted by (\ref{h1}).

\section*{Linear hidden central charges}

We assume now that (\ref{L}) is part
of a globally supersymmetric action whose
remaining part contains only matter fields and is
quadratic in these fields. Accordingly,
the supersymmetry transformations are assumed to
be linear in the fields. One should have in mind here
in particular models in a flat spacetime
but the arguments are not restricted to that case.
The linearity of the supersymmetry transformations
implies that the 
hidden symmetries which can possibly occur
in the commutators of supersymmetry transformations
leave the matter fields invariant and
act on the gauge fields according to
\beq
\delta_{\mathrm{hidden}}^{\mathrm{linear}} 
A^a= \sum c^{ab} \star dA^b\ ,\quad
c^{ab}=(-)^{np_a}c^{ba},
\label{A}
\eeq
where the sum 
runs over all values of $b$ such that
$p_a=n-1-p_b$. Furthermore, the
transformations (\ref{A}), and in fact
all other transformations (\ref{h1}) as well,
commute on-shell with the supersymmetry transformations,
\beq
[\delta_{\mathrm{hidden}},\delta_{\mathrm{susy}}]\approx 0.
\label{B}
\eeq
Obviously they
also commute with the spacetime symmetries
generated by Lie derivatives of the fields
along Killing vector fields of the background
metric.

These statements hold on simple and general grounds.
Let us denote
the commutator algebra of supersymmetry transformations
on the field strengths and on the matter fields
by $[\delta_{\mathrm{susy}},\delta'_{\mathrm{susy}}]
\approx \delta_{\mathrm{comm}}$ where
$\delta_{\mathrm{comm}}$ is some linear symmetry transformation.
This gives
$d([\delta_{\mathrm{susy}},\delta'_{\mathrm{susy}}]A^a- 
\delta_{\mathrm{comm}}A^a)\approx 0$. Using the results on the
characteristic cohomology described above and the
fact that $[\delta_{\mathrm{susy}},\delta'_{\mathrm{susy}}]
-\delta_{\mathrm{comm}}$
is a linear symmetry, one concludes
$[\delta_{\mathrm{susy}},\delta'_{\mathrm{susy}}]A^a- 
\delta_{\mathrm{comm}}A^a \approx
\delta_{\mathrm{hidden}}^{\mathrm{linear}}A^a+dY^a$
for some local forms $Y^a$ which are linear in the fields
and may depend explicitly on the $x^\mu$ (see below).
Hence, one has indeed
$[\delta_{\mathrm{susy}},\delta'_{\mathrm{susy}}]\approx
\delta_{\mathrm{comm}}+\delta_{\mathrm{hidden}}^{\mathrm{linear}}
+\delta_{\mathrm{gauge}}(Y)$ with  
$\delta_{\mathrm{hidden}}^{\mathrm{linear}}$ as described above.

(\ref{B}) holds because linear supersymmetry 
transformations commute
with the gauge transformations (\ref{gsymm}),
\beq
[\delta_{\mathrm{susy}},\delta_{\mathrm{gauge}}(\epsilon)]=0,
\label{c1}
\eeq  
where the gauge parameter fields are inert to
supersymmetry transformations, 
$\delta_{\mathrm{susy}}\epsilon^a_{\mu_1\dots\mu_{p_a-1}}=0$.
Indeed,
the commutator of any global symmetry transformation
and a general gauge transformation (with arbitrary parameter fields
inert to the global symmetry)
either vanishes on-shell or generates
a nontrivial gauge symmetry of the action on-shell.
In our case
$[\delta_{\mathrm{susy}},\delta_{\mathrm{gauge}}(\epsilon)]$ is
field independent when evaluated on any of the fields
($\delta_{\mathrm{susy}}$ is linear in the fields, while
the gauge transformations (\ref{gsymm}) do not involve the
fields).
Hence, the field equations
cannot appear in
$[\delta_{\mathrm{susy}},\delta_{\mathrm{gauge}}(\epsilon)]$
because they are linear in the fields.
Furthermore 
$[\delta_{\mathrm{susy}},\delta_{\mathrm{gauge}}(\epsilon)]$
is not a gauge transformation
(\ref{gsymm})  as one has
$[\delta_{\mathrm{susy}},\delta_{\mathrm{gauge}}(\epsilon)]A^a=0$
($\delta_{\mathrm{susy}}\delta_{\mathrm{gauge}}(\epsilon)A^a=0$
owing to $\delta_{\mathrm{susy}}\epsilon^a_{\mu_1\dots\mu_{p_a-1}}=0$,
and $\delta_{\mathrm{gauge}}(\epsilon)\delta_{\mathrm{susy}}A^a=0$
because $\delta_{\mathrm{susy}}A^a$ is a linear combination of fermionic 
fields).
This gives (\ref{c1}) and implies that
a supersymmetry transformation of any field is gauge
invariant and thus that the
gauge fields $A^a_{\mu_1\dots\mu_{p_a}}$ can enter
the supersymmetry transformations only via
the field strengths $F^a_{\mu_0\dots\mu_{p_a}}$.
This implies (\ref{B}) because of $\delta_{\mathrm{hidden}} 
F^a_{\mu_0\dots\mu_{p_a}}\approx 0$.

Note that it only
depends on the field content and the spacetime dimension
whether or not there is
a hidden symmetry (\ref{A}):
there must be at least two
gauge potentials whose form-degrees add up to $(n-1)$
(these two gauge potentials
may coincide if $n=4k+1$ and $p_a=2k$). 

However, the fact that the hidden symmetries commute with 
the spacetime symmetries
limits the possible situations in
which they can be present in the commutators
of supersymmetry transformations. For instance, 
Poincar\'e-invariant models in flat four-dimensional spacetime
must have extended ($N\geq 2$) supersymmetry in order that
this can happen (see below).

\section*{Remark on $x$-dependence}

It should be stressed that all statements about the characteristic
cohomology and hidden symmetries made above
refer to the space of local forms and transformations which are allowed
to depend explicitly on the spacetime coordinates $x^\mu$.
This is relevant even in a flat background. Indeed, 
for an action (\ref{L}), the characteristic
cohomology at form-degrees $<n-1$ is bigger in the
restricted space of $x$-independent local forms than in the
space of all local forms. The additional
representatives are exterior products of the
$\star dA^a$ and at least one differential $dx^\mu$ \cite{HKS}.
They give rise to symmetries of
the action analogous to (\ref{h1}), but, in contrast
to the latter, these symmetries are equal to gauge
transformations on-shell
and are thus trivial global symmetries according
to modern terminology (cf.\ \cite{report}, section 6).

Of particular importance in the context of linear supersymmetry are the 
linear symmetries of this type because they can show up
in the commutators of supersymmetry
transformations (see \cite{TT} and the example below).
They are given by
\bea
\delta_{\mathrm{trivial}} A^a
&=& \sum c^{ab}_p\wedge (\star dA^b),
\nonumber\\
c^{ab}_p&=&(-)^{(n-p_a-1)(n-p_b-1)}c^{ba}_p\ ,\quad p>0
\label{C}
\eea
where the sum runs over values of $b$ and $p$
such that $p_a=p+n-1-p_b$, and
$c^{ab}_p$ are $p$-forms with constant coefficients,
\[
c^{ab}_p=\frac 1{p!}c^{ab}_{\mu_1\dots\mu_p}dx^{\mu_1}\wedge\dots 
\wedge dx^{\mu_p},\quad c^{ab}_{\mu_1\dots\mu_p}=\mathrm{constant}.
\]
The $x$-dependent gauge transformations corresponding to
(\ref{C}) are easily found using $dx^\mu=d(x^\mu)$,
\bea
&&\delta_{\mathrm{trivial}} A^a\approx dX^a\ \Leftrightarrow\ 
\delta_{\mathrm{trivial}} A^a\approx
\delta_{\mathrm{gauge}}(X)A^a,
\nonumber\\
&&X^a=\sum
\frac 1{p!}c^{ab}_{\mu_1\dots\mu_p}x^{\mu_1} dx^{\mu_2}\wedge\dots 
\wedge dx^{\mu_p}\wedge \star dA^b.
\label{D}
\eea

\section*{Examples in 4 dimensions}

Examples in 4-dimensional spacetime
involve only 1-form or 2-form gauge potentials.
Symmetries (\ref{A}) then shift 1-form (2-form) gauge potentials
by the Hodge dualized field strengths of 2-form (1-form) gauge
potentials. Hence, models with such symmetries must
contain both at least one 1-form gauge potential and
at least one 2-form gauge potential. 

Furthermore, Poincar\'e invariant models in
flat 4-dimensional spacetime
must have extended ($N\geq 2$) supersymmetry
in order that the commutator of two supersymmetry
transformations can contain a symmetry (\ref{A}).
This is seen when one writes the
supersymmetry transformations as
\[
\delta_{\mathrm{susy}}=\sum_{i=1}^N (\xi^{\alpha i}D_\alpha^i
+\bar \xi_{\dot\alpha}^i\bar D^{{\dot\alpha} i})
\]
where $\xi^{\alpha i}$ are constant anticommuting Weyl-spinors,
$\bar \xi^{{\dot\alpha} i}$ are their complex conjugates,
and $D_\alpha^i$ and $\bar D_{{\dot\alpha}}^i$ generate
the corresponding supersymmetry transformations of the fields
(using conventions as in \cite{WB} for 
the Minkowski metric $\mathrm{diag}(1,-1,-1,-1)$).
The commutator of two supersymmetry transformations
involves the anticommutators $\{D_\alpha^i,\bar D_{\dot\alpha}^j\}$
and $\{D_\alpha^i,D_\beta^j\}$ (and the
complex conjugates of the latter). $\{D_\alpha^i,\bar D_{\dot\alpha}^j\}$
contains no Lorentz-invariant piece and therefore
it cannot contain a hidden symmetry (\ref{A}). In contrast,
$\{D_\alpha^i,D_\beta^j\}$
can contain
$\varepsilon_{\alpha\beta}\delta_{\mathrm{hidden}}^{ij}$
where $\delta_{\mathrm{hidden}}^{ij}=-\delta_{\mathrm{hidden}}^{ji}$ 
are hidden symmetries (\ref{A}). The antisymmetry in $i,j$ requires
$N\geq 2$.
 
The VT multiplet \cite{VT} meets these conditions.
It contains one 1-form gauge potential, one 2-form gauge potential,
one real scalar field and two Weyl fermions (one may add
an auxiliary real scalar field).
In that case one has $\{D_\alpha^i,D_\beta^j\}
\approx \varepsilon_{\alpha\beta}\varepsilon^{ij}\delta_{\mathrm{hidden}}$
with $\delta_{\mathrm{hidden}}$ as in (\ref{A}).
Another example, the VTT-multiplet, will be given below.

Trivial symmetries (\ref{C}) in four dimensions act nontrivially only
on 2-form gauge potentials and involve constant 1-forms
$c_1^{ab}=-c_1^{ba}$.  The antisymmetry
in $a,b$ requires the presence of at least two 2-form
gauge potentials. As the $c_1^{ab}$ are 1-forms, these
symmetries are not Lorentz-scalars but Lorentz-vectors. 
Therefore they do not
occur in $\{D_\alpha^i,D_\beta^j\}$ but
in $\{D_\alpha^i,\bar D_{\dot\alpha}^j\}$ and this can happen
already for $N=1$ supersymmetry.

A multiplet where a symmetry (\ref{C})
occurs in the commutator of supersymmetry transformations 
is the $N=2$ double tensor
multiplet \cite{TT}. Another example with $N=2$
supersymmetry is the VTT multiplet. An example with $N=1$ supersymmetry
is the ``$N=1$ double tensor multiplet'' considered in \cite{PvN}. 
It can be obtained
by truncating the VTT multiplet, see below.

The VTT multiplet contains
one 1-form gauge potential $A=dx^\mu A_\mu$, two
2-form gauge potentials $B^a=(1/2)dx^\mu\wedge dx^\mu B^a_{\mu\nu}$ 
($a=1,2$) and
two Weyl fermions $\psi^i_\alpha$ ($i=1,2$).
It is convenient to combine the 2-form gauge fields
in complex fields, 
\[
B=\frac 12\,dx^\mu\wedge dx^\mu B_{\mu\nu}\ ,\quad
B_{\mu\nu}=B^1_{\mu\nu}+{\mathrm{i}} B^2_{\mu\nu}\ .
\] 
The Lagrangian for the free multiplet is
\[
L=-\frac 14\, F_{\mu\nu}F^{\mu\nu}-\frac 12\, H_\mu \bar H^\mu
-2{\mathrm{i}}\, \psi^i\partial \bar \psi^i
\]
where
\beann
F_{\mu\nu}&=&\partial_\mu A_\nu-\partial_\nu A_\mu\ ,
\\
H^\mu&=&\frac 16\,\varepsilon^{\mu\nu\rho\sigma}F_{\nu\rho\sigma}
=\frac 12\,\varepsilon^{\mu\nu\rho\sigma}\partial_\nu B_{\rho\sigma}\ .
\eeann
The N=2 supersymmetry transformations $D_\alpha^i$ to be discussed read
\beann
&D_\alpha^i A_\mu = \varepsilon^{ij}(\sigma_\mu \bar \psi^j)_\alpha\ ,&
\\ 
&D_\alpha^i B_{\mu\nu} = 4 (\sigma_{\mu\nu} \psi^i)_\alpha\ ,\ 
D_\alpha^i \bar B_{\mu\nu} = 0\ ,& \\
&D_\alpha^i\psi^j_\beta = -\frac{{\mathrm{i}}}2 \varepsilon^{ij}
                           \sigma^{\mu\nu}_{\alpha\beta}F_{\mu\nu}\ ,\ 
D_\alpha^i\bar \psi_{\dot\alpha}^j = 
-\frac 12 \delta^{ij}\, \bar H_{\alpha{\dot\alpha}}\ .&
\eeann
The $\bar D_{\dot\alpha}^i$ are obtained by complex conjugation.
The anticommutators $\{D_\alpha^i,\bar D_{\dot\alpha}^j\}$ read
\bea
\{D_\alpha^i,\bar D_{\dot\alpha}^j\}\psi^k_\beta & = &
   -{\mathrm{i}}\, \delta^{ij}\,\partial_{\alpha{\dot\alpha}}\psi^k_\beta 
\nonumber\\
   &&+\frac{{\mathrm{i}}}2\, \varepsilon_{\alpha\beta}\, 
   (3\delta^{jk}\delta^{il}-\delta^{ij}\delta^{kl})
   \partial_{\gamma{\dot\alpha}}\psi^{\gamma l}
\nonumber\\
   &\approx& -{\mathrm{i}}\, \delta^{ij}\,
             \partial_{\alpha{\dot\alpha}}\psi^k_\beta 
\nonumber\\
\{D_\alpha^i,\bar D_{\dot\alpha}^j\}A_\mu &= & 
   -{\mathrm{i}}\, \delta^{ij}\, \sigma^\nu_{\alpha{\dot\alpha}}\, F_{\nu\mu}
\nonumber\\
\{D_\alpha^i,\bar D_{\dot\alpha}^j\}B_{\mu\nu} &= & 
   -{\mathrm{i}}\, \delta^{ij}\, \sigma^\rho_{\alpha{\dot\alpha}} 
   F_{\rho\mu\nu}
\nonumber\\
   & &+\delta^{ij}\,
   (H_\nu\sigma_\mu-H_\mu\sigma_\nu)_{\alpha{\dot\alpha}}\ .
\label{alg1}
\eea
The terms $H_\nu\sigma_\mu-H_\mu\sigma_\nu$ 
in $\{D_\alpha^i,\bar D_{\dot\alpha}^j\}B_{\mu\nu}$
make up a symmetry (\ref{C}) and 
are thus $x$-dependent gauge transformations
of $B^1_{\mu\nu}$ and $B^2_{\mu\nu}$ on-shell (see below). 
$F_{\nu\mu}$ and $F_{\rho\mu\nu}$ are 
modulo particular gauge transformations of
$A_\mu$ and $B_{\mu\nu}$ equal to 
$\partial_\nu A_\mu$ and $\partial_\rho B_{\mu\nu}$ respectively.
Hence, $\{D_\alpha^i,\bar D_{\dot\alpha}^j\}$
equals $-{\mathrm{i}}\, \delta^{ij}\,\partial_{\alpha{\dot\alpha}}$ 
plus gauge transformations on-shell.

The anticommutators $\{D_\alpha^i,D_\beta^j\}$ read
\bea
\{D_\alpha^i,D_\beta^j\}\psi^k_\gamma& = &
   {\mathrm{i}} \varepsilon^{jk}\varepsilon^{il}
   \varepsilon_{\alpha(\beta}\partial_{\gamma){\dot\alpha}}
   \bar \psi^{{\dot\alpha} l}
   +(\alpha i\leftrightarrow \beta j)
   \approx  0
\nonumber\\
\{D_\alpha^i,D_\beta^j\}\bar \psi_{{\dot\alpha} k}& = & 0
\nonumber\\
\{D_\alpha^i,D_\beta^j\} A_\mu
   &=&-\varepsilon^{ij}\varepsilon_{\alpha\beta}\, \bar H_\mu
\nonumber\\
\{D_\alpha^i,D_\beta^j\}B_{\mu\nu} 
   &=&-\varepsilon^{ij}\varepsilon_{\alpha\beta}
   (\varepsilon_{\mu\nu\rho\sigma}F^{\rho\sigma}
   +2{\mathrm{i}} F_{\mu\nu})
\nonumber\\
\{D_\alpha^i,D_\beta^j\}\bar B_{\mu\nu} &=& 0.
\label{alg2}\eea
The second term in parantheses in
$\{D_\alpha^i,D_\beta^j\}B_{\mu\nu}$ 
is a gauge transformation. Except for this gauge transformation,
$\{D_\alpha^i,D_\beta^j\}$ is given by
$-\varepsilon^{ij}\varepsilon_{\alpha\beta}
(\delta^1_{\mathrm{hidden}}-{\mathrm{i}} \delta^2_{\mathrm{hidden}})$
on-shell and contains two
hidden symmetries:
\bea
&&\delta^a_{\mathrm{hidden}}A_\mu=H_\mu^a\ ,\ 
\delta^a_{\mathrm{hidden}}B^b_{\mu\nu}=\frac 12\delta^{ab}
\varepsilon_{\mu\nu\rho\sigma}F^{\rho\sigma}\ ,
\nonumber\\
&& \delta^a_{\mathrm{hidden}}\psi^i_\alpha=0
\quad (a,b=1,2).
\label{alg3}
\eea

Hence, the commutator of two supersymmetry transformations
is given by the sum of a translation, two hidden symmetries (\ref{A})
and gauge transformations on-shell,
\bea
[\delta_{\mathrm{susy}},\delta'_{\mathrm{susy}}]&\approx&
-a^\mu \partial_\mu+(a+\bar a)\delta^1_{\mathrm{hidden}}
-{\mathrm{i}}(a-\bar a)\delta^2_{\mathrm{hidden}}
\nonumber\\
& &+\delta_{\mathrm{gauge}}(\epsilon_\mu=
a^\nu B_{\nu\mu}-{\mathrm{i}} a_\nu x^\nu H_\mu+2{\mathrm{i}} a A_\mu)
\nonumber\\
&& +\delta_{\mathrm{gauge}}(\epsilon=a^\mu A_\mu),
\nonumber\\
a^\mu&=&{\mathrm{i}} (\xi^i\sigma^\mu\bar \xi^{i\prime}
-\xi^{i\prime}\sigma^\mu\bar \xi^i),\quad
a=\varepsilon^{ij}\xi^i\xi^{j\prime}
\label{alg4}
\eea
where $\delta_{\mathrm{gauge}}(\epsilon=\dots)$ and 
$\delta_{\mathrm{gauge}}(\epsilon_\mu=\dots)$
are gauge transformations of $A_\mu$ and
$B_{\mu\nu}$ with particular gauge parameter fields respectively
($\epsilon_\mu=\epsilon^1_\mu+{\mathrm{i}}\epsilon^2_\mu$).
Notice that the $x$-dependent
gauge transformations in
$[\delta_{\mathrm{susy}},\delta'_{\mathrm{susy}}]$ arise indeed
from a symmetry (\ref{C}) given by
\bea
\delta_{\mathrm{trivial}} B&=&-\frac{{\mathrm{i}}}{2}\,
dx^\mu a_\mu\wedge dx^\nu H_\nu
\nonumber\\ 
\Leftrightarrow\ 
\delta_{\mathrm{trivial}} B^a&=&\frac 12\,\varepsilon^{ab}dx^\mu a_\mu
\wedge \star dB^b.
\label{alg5}
\eea

If one sets $A_\mu=0$, $\psi^2_\alpha=0$
and drops $D_\alpha^2$,  the above formulae
provide the Lagrangian, the supersymmetry
transformations and the commutator algebra of these transformations
for the $N=1$ double tensor multiplet \cite{PvN}.
The symmetry (\ref{alg5}) appears then
in the $N=1$ commutators 
$\{D_\alpha^1,\bar D_{\dot\alpha}^1\}$.

\section*{Remark on dualities}

Assume that two linear supersymmetric models of the type considered above
are related by a standard duality substituting an $(n-p-2)$-form
gauge potential $\tilde A$
for a $p$-form gauge potential $A$ (including the
case $p=0$ representing scalar fields).
More precisely, the duality substitutes $\star d\tilde A$
for $dA$ both in the Lagrangian and in the supersymmetry transformations
(with appropriate coefficients), and thus field equations for
Bianchi identities (and vice versa). Hence,
if an expression in the field strength components
and their derivatives
vanishes in the original model then its counterpart in
the dual model vanishes at least on-shell.

As a consequence, the duality does not
modify the on-shell supersymmetry algebra
on the field strengths and matter fields 
(recall that the gauge fields occur in the
supersymmetry transformations only via the field strengths).
The only effect of the duality
on the on-shell algebra of supersymmetry transformations
is thus a possible modification of the hidden symmetries and the gauge
transformations which appear in the algebra.
In particular, hidden symmetries (\ref{A}) and/or
trivial symmetries (\ref{C}) may thus be present in the
supersymmetry algebra after a duality transformation even when
they were absent before.

For instance, the hidden central charge symmetry of the
VT multiplet arises in this way by dualizing
one of the two real scalar fields of an abelian $N=2$ vector
multiplet. Similarly, the two hidden central charge symmetries
(\ref{alg3}) of the VTT multiplet arise by dualizing both 
scalar fields of 
an abelian $N=2$ vector multiplet. The latter duality
gives also rise to
the symmetry (\ref{alg5}) in the commutators of
supersymmetry transformations.
Analogous statements apply to the $N=2$ double tensor 
multiplet \cite{TT} and the $N=1$
double tensor multiplet \cite{PvN} as these multiplets are
dual to an
$N=2$ hyper multiplet and an $N=1$ chiral multiplet respectively.

\section*{Consequences for interactions}

Even though the previous discussions apply to
linear models and supersymmetry transformations, they are 
also relevant to nonlinear supersymmetric extensions of such models.
In particular they matter to the construction of
interacting supersymmetric models from free ones.
For instance, suppose that one looks for a nonlinear extension
of a free supersymmetric model 
with a hidden symmetry
occurring in the commutators of 
supersymmetry transformations. Then this hidden symmetry must
have a counterpart in the algebra of supersymmetry transformations
of the nonlinear model, i.e., the nonlinear
model must have a symmetry of a similar type.
In general this symmetry
will be a nonlinear extension of the corresponding
hidden symmetry of the free model.
This can restrict the possible interactions
quite severely. 
There are two possibilities to be distinguished:

(i) The hidden symmetry of the free model
is extended to a nontrivial global symmetry of the
interacting model. Examples have been constructed in 
\cite{glVT1,glVT2,glVT3,glVT4,glVT5}
where the central charge symmetry of the
free VT multiplet is extended to a nonlinear
global symmetry.

(ii) The hidden symmetry of the free model is promoted to
a gauge symmetry of the interacting model.
This case is very interesting. In particular,
the gauging of central charge symmetries is inevitable
when one wants to couple a supersymmetry multiplet 
with such a symmetry to supergravity.  
Four-dimensional globally supersymmetric models with a gauged
hidden central charge were constructed
in \cite{VTgauged1,VTgauged2,VTgauged3,VTgauged4,T}, 
again for the VT multiplet.
Supergravity models with VT multiplets were constructed
in \cite{sugraVT1} (see also \cite{sugraVT2}). 

In fact, the gauging of hidden symmetries is
interesting in its own right, whether or not these symmetries
occur in the commutator of supersymmetry transformations.
It is a nontrivial matter already
in the non-supersymmetric case, as was discussed
for a particular example in \cite{BD}
and in greater generality in \cite{BST}.
Supersymmetry makes gauging of hidden symmetries
even more involved because it requires to combine
it with other interactions
between $p$-form gauge fields and matter fields
(in some cases it may even be impossible
to gauge a hidden symmetry in a supersymmetric way).

The gauging of hidden symmetries
may be combined with other interactions peculiar to
$p$-form gauge potentials \cite{HK}, such as 
Freedman-Townsend interactions \cite{OP,FT} or couplings 
to Chern-Simons-forms. 
A large class of such supersymmetric models
in four dimensions 
was constructed in \cite{BT1}
(see also \cite{BT2}).
Among others it was found there
that gauged hidden symmetries in general do not
commute with the supersymmetry transformations on-shell,
in contrast to their counterparts in free models (cf.\ Eq.\ (\ref{B})).
Analogous models in dimensions $n>4$ might be 
interesting especially in the string theory context
but it seems that no such models have been constructed yet.

\end{document}